\documentclass[french,english,pra,twocolumn]{revtex4}
\usepackage[T1]{fontenc}
\usepackage[latin9]{inputenc}
\usepackage{array}
\usepackage{multirow}
\usepackage{amsmath}
\usepackage{amssymb}
\usepackage{graphicx}

\makeatletter

\providecommand{\tabularnewline}{\\}

\@ifundefined{textcolor}{}
{%
 \definecolor{BLACK}{gray}{0}
 \definecolor{WHITE}{gray}{1}
 \definecolor{RED}{rgb}{1,0,0}
 \definecolor{GREEN}{rgb}{0,1,0}
 \definecolor{BLUE}{rgb}{0,0,1}
 \definecolor{CYAN}{cmyk}{1,0,0,0}
 \definecolor{MAGENTA}{cmyk}{0,1,0,0}
 \definecolor{YELLOW}{cmyk}{0,0,1,0}
}

\@ifundefined{showcaptionsetup}{}{%
 \PassOptionsToPackage{caption=false}{subfig}}
\usepackage{subfig}
\makeatother

\usepackage{babel}
\makeatletter
\addto\extrasfrench{%
   \providecommand{\fg}{\ifdim\lastskip>\z@\unskip\fi~\frqq}%
}

\makeatother
\begin{document}

\title{Measuring the re-absorption cross section of a Magneto-Optical Trap}

\author{Rudy Romain, Hélène Louis, Philippe Verkerk and Daniel Hennequin}

\address{Laboratoire PhLAM, CNRS UMR 8523, Bât. P5 - Université Lille1, 59655
Villeneuve d'Ascq cedex, FRANCE}

\email{daniel.hennequin@univ-lille1.fr}

\begin{abstract}
Magneto-Optical Traps have been used for several decades. Among fundamental
mechanisms occuring in such traps, the magnitude of the multiple scattering
is still unclear. Indeed, many experimental situations cannot be modeled
easily, different models predict different values of the re-absortion
cross section, and no simple experimental measurement of this cross
section are available. We propose in this paper a simple measurement
of this cross section through the size and the shape of the cloud
of cold atoms. We apply this method to traps with a configuration
where theoretical values are available, and show that the measured
values are compatible with some models. We also apply the method to
configurations where models are not relevant, and show that the re-absorption
is sometimes much larger than the usually assumed value. 
\end{abstract}
\maketitle

\section{Introduction}

The Magneto-Optical Trap (MOT) is today an essential tool in experimental
atomic physics. A MOT produces ultracold atoms which can be used in
numerous experiments: MOTs are used -- alone or with further cooling
steps -- to study quantum chaos \cite{ghose2008}, Anderson localization
\cite{jendrzejewski2012} or plasma instabilities \cite{tercas2013,romain2011},
to produce ultra precise atomic clocks \cite{hinkley2013}, to obtain
cold molecules \cite{molecules}, and many other studies.

In most experiments, an accurate knowledge of the detailed physical
mechanisms occurring in a MOT is not necessary. For example, it is
well known that the atomic cloud produced by a MOT becomes unstable
for some parameters, but it is also well known that in most cases,
the alignment of the laser beams has just to be adjusted to remove
these instabilities. However, recent studies have shown that understanding
the detailed physics of the MOT could be interesting, because the
MOT appears to belong to the large family of complex systems described
by Vlasov-Fokker-Planck (VFP) equations \cite{romain2011}. Because
MOT experiments are relatively simple to implement within this family,
they appear as a good candidate for a model system in this area. And
anyway, understanding in detail the physics of MOT would allow to
understand what is observed in the experiments.

The mechanisms allowing to obtain a cloud of cold atoms with a MOT
are well known. They result from the interactions of the atoms with
the laser light. The basic interaction is the absorption of a photon
by an atom, followed by a spontaneous emission process. Depending
in particular on the cloud density, the emitted photon can possibly
undergo several cycles of absorption/emission before escaping the
cloud: it is the multiple scattering. The final cloud of cold atoms
is the result of the equilibrium between the trapping force (induced
by the exchange of momentum during the diffusion cycle), the shadow
effect force (induced by the absorption of the beam through the cloud),
and the multiple scattering force. The amplitude of the forces depends
in particular on the absorption cross section $\sigma_{L}$ and on
the re-absorption cross section $\sigma_{R}$. Knowing these quantities
is thus essential to model accurately a MOT. Moreover these cross
sections play a key role in the analogy with plasma physics because
the effective charge involved in the coulomb-like interaction between
cold atoms depends on their ratio \cite{sesko1991}.

The detailed mechanisms involved in a MOT are rather complex. Taking
into account the exact distribution of the atomic levels and the interaction
of the atoms with all the laser beams would lead to a very complex
set of equations. Thus all the existing models deal with approximations
which lead to different predictions of the cross section values. This
is not a big deal concerning $\sigma_{L}$ because simple absorption
measurements lead to this quantity. On the contrary, no simple way
to measure $\sigma_{R}$ has been proposed until now. So theoretical
predictions have never been validated by experimental measurements,
and the $\sigma_{R}$ values found today in the literature are still
questionable.

We propose in this paper a method to measure the re-absorption cross
section in a magneto-optical trap. We apply this method to a Cesium
trap and obtain values of $\sigma_{R}$ for different sets of laser
parameters. We compare these values to different models.

\section{Theory}

We consider here a MOT in the usual $\sigma^{+}-\sigma^{-}$ configuration.
Each of the three arms of the trap consists in a pair of counter-propagating
laser beams characterized by their intensity and their frequency.
A pair is obtained by retro-reflection of an incident beam. The beam
intensities are respectively $I_{+}$ for the incident beam and $I_{-}$
for the retro-reflected one, while their frequency is given through
the detuning $\Delta$ between the laser frequency and the atomic
transition. We do not need to detail the energetic structure of the
cold atoms in order to understand the steady state of the cloud. Indeed,
we adopt a global formalism in term of cross sections.

The stationary cloud of cold atoms results from the equilibrium between
three forces acting on the atoms. These forces have been expressed
in numerous studies, with more or less approximations \cite{sesko1991,steane1992,townsend1995,pruvost2000,romain2011}.
However, all these models deal with the same variables and the same
parameters. The first force is the trapping force produced by the
laser beams and the Zeeman shifts induced by the magnetic field. It
is a restoring force, characterized by the spring constant $\kappa$.
The two other forces are collective forces. The shadow effect force
is due to the absorption of the laser beams all along the cloud, leading
to a local imbalance of the laser beam intensities. Thus this force
depends on the absorption cross section $\sigma_{L}$. Finally, the
multiple scattering force is induced by additional scattering of photons,
and so depends on the re-absorption cross section $\sigma_{R}$. The
first two forces compress the cloud of atoms, while the latter causes
it to expand. The size of the obtained atomic cloud is thus the result
of an equilibrium between these three forces.

To study the equilibrium, we use an approach similar to that in \cite{sesko1991}
which assumes that the temperature of the cloud is zero. The main
assumption in this model is that a photon is re-scattered at most
once before escaping the cloud. Contrary to \cite{sesko1991}, we
consider the usual anti-Helmholtz configuration for the coils creating
the magnetic field, as mentioned previously. Such a field is zero
at the point defined as the center of the trap and it can be assumed
to be linear along each direction. We take into account the fact that
the magnetic field gradient along the coil axis is twice that along
the perpendicular directions. So, the spherical symmetry used in \cite{sesko1991}
is broken, and the cloud shape must be modeled as an ellipsoid.

The determination of the stationary density $n$ does not need the
knowledge of the forces. Indeed, the collective forces depend on the
shape of the cloud, while their divergences do not. The vanishing
of the divergence of the total force (which is zero at equilibrium)
gives us a constant atomic density. The results from \cite{romain2013},
where a more general anisotropic configuration has been considered,
can be applied to the present situation and give: 
\begin{equation}
n=\frac{2c\kappa}{3I_{+}\,\sigma_{L}^{2}\left(S-1\right)}\label{eq:densite isotrope}
\end{equation}
where $S=\sigma_{R}/\sigma_{L}$ is the cross section ratio and $c$
is the speed of light. Note that as the quadrupolar magnetic field
is taken into account, this expression differs slightly from that
in \cite{sesko1991,steane1992}. We can improve this description of
the MOT by calculating the expression of the three forces. The trapping
force has a usual form and takes into account the anisotropy of the
magnetic field gradient. The shadow effect force has the same expression
as is \cite{sesko1991}, the absorption is assumed to be linear. The
net multiple scattering force is calculated as the sum over the cloud
of all the coulomb-like atom-atom interactions associated with a scattering
process. In \cite{romain2013}, it is shown how to calculate this
force for an ellipsoidal cloud, with half-widths $L_{\parallel}$
and $L_{\perp}$ along the coil axis and in the transverse plane respectively.
A geometric parameter $A$, depending only on the ellipticity $\varepsilon=L_{\perp}/L_{\parallel}$
of the cloud, appears in the expression of the force:
\begin{eqnarray}
A & = & \frac{\varepsilon^{2}}{\varepsilon^{2}-1}\,\beta\label{eq:A=00003Df(ellip)}\\
{\displaystyle \beta} & = & {\displaystyle \begin{cases}
{\displaystyle 1-\frac{1}{\sqrt{1-\varepsilon^{2}}}\ln\left|\frac{1+\sqrt{1-\varepsilon^{2}}}{1-\sqrt{1-\varepsilon^{2}}}\right|} & for\:\varepsilon^{2}>1\\
1-\frac{1}{\sqrt{{\displaystyle \varepsilon^{2}-1}}}\arcsin\left(\sqrt{{\displaystyle \frac{\varepsilon^{2}-1}{\varepsilon^{2}}}}\right) & for\:\varepsilon^{2}<1
\end{cases}}\nonumber 
\end{eqnarray}
The equilibrium of the forces gives another condition on this parameter
:
\begin{equation}
{\displaystyle A=\frac{1}{2}\left(1-{\displaystyle \frac{1}{3S}}\right)}\label{eq:A=00003Df(sigma)}
\end{equation}
For the sake of completeness, the equilibrium of the forces also gives
us access to the cloud displacement due to the shadow effect in our
asymmetric configuration. This displacement is global, and depends
only on the shadow effect: it gives no information about multiple
scattering.

From the previous equations, it is easy to find that $\varepsilon$
is bounded. First, the atomic density is a positive quantity so that,
in eq. \ref{eq:densite isotrope}, $\sigma_{R}$ has to be larger
than $\sigma_{L}$ ($S>1)$. It follows from eq. \ref{eq:A=00003Df(sigma)}
that $A>1/3$. From the same equation, we also find that the maximum
value of $A$ is $1/2$ (when $S\to+\infty$). This leads to:

\begin{equation}
1<\varepsilon<1.81
\end{equation}
 Thus, the atomic cloud shape of a usual MOT appears to be always
oblate. A spherical cloud corresponds to a infinite density, and so
is not a physical solution. We also point out an upper limit to the
ellipticity which corresponds to $\sigma_{R}\gg\sigma_{L}$. 

\selectlanguage{french}%
\begin{figure}[t]
\selectlanguage{english}%
\centering{}\includegraphics[width=8.6cm]{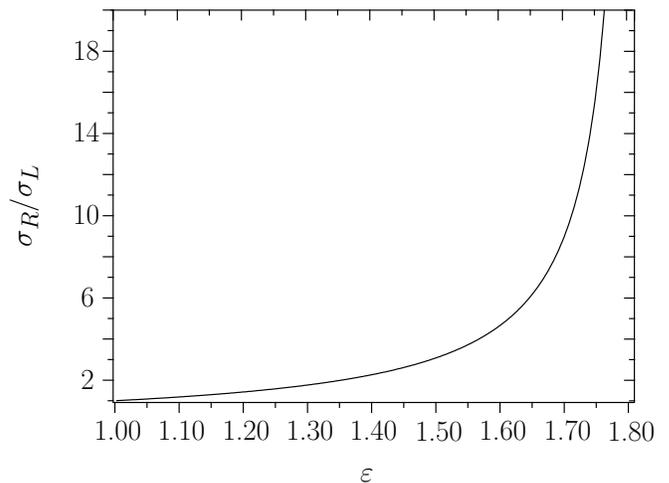}\foreignlanguage{french}{\caption{\selectlanguage{english}%
Evolution of the cross section ratio inside the predicted range of
ellipticity values\foreignlanguage{french}{.\label{fig:sect eff=00003Df(ellip)}}\selectlanguage{french}%
}
}\selectlanguage{french}%
\end{figure}

\selectlanguage{english}%
Fig. \ref{fig:sect eff=00003Df(ellip)} shows the cross section ratio
versus the ellipticity for all possible values: the more elongated
the cloud, the more probable the re-absorption. This result is quite
different from the one predicted for the temperature-limited regime
in which the ellipticity of the cloud is constant and equal to $\sqrt{2}$
\cite{townsend1995}. In the multiple scattering regime, the ratio
$S$ and the cloud ellipticity depend \emph{a priori} on the laser
parameters.

As pointed out above, the calculation of the re-absorption cross section
as a function of the laser parameters is rather complex. The atom
must be modeled with a particular atomic structure from which the
overlap between the emission and the absorption spectra of a cold
atom can be calculated. Several authors figured out theoretical expressions
of $\sigma_{R}$ in the context of Doppler cooling theory \cite{sesko1991,steane1992,townsend1995,pruvost2000,romain2011},
but with different approaches. In \cite{sesko1991}, a fit of single
cloud size measurements is performed for only one set of parameter
values. \cite{steane1992} performs an approximated analytical calculation,
while \cite{townsend1995,pruvost2000,romain2011} use the dressed
atom picture (DAP) \cite{cohen}. All consider a two-level atom, except
\cite{romain2011} who considers a three-level atom. These calculations
take into account the saturation by the two counter-propagating beams,
requiring to introduce the total Rabi frequency $\Omega=\Gamma\sqrt{\left(I_{+}+I_{-}\right)/2I_{sat}}$,
with $I_{sat}$ the saturation intensity ($I_{sat}=1.1$ mW/cm$^{2}$
for Cs) and $\Gamma$ the natural width of the transition.

Within the DAP, the secular approximation limits the range of parameter
values to $\Omega^{2}+\Delta^{2}\gg\Gamma^{2}$. But even with such
a simplification, the expression of $\sigma_{R}$ remains heavy. Table
\ref{tab:compa sect eff} summarizes the expressions obtained by these
previous studies, in the two limit cases where the laser intensity
is much larger than the detuning and vice versa. On the one hand,
these two situations ($\left|\Delta\right|\gg\Omega\gg\Gamma$ and
$\Omega\gg\left|\Delta\right|\gg\Gamma$) give simple asymptotic expressions.
And on the other hand, they represent usual experimental parameters.
Note that the values of parameters in \cite{sesko1991} do not fit
these two limit cases, and the results from \cite{steane1992} do
not seem to be relevant. The expressions derived from the DAP give
the same dependence on the laser parameters, but with different numerical
factors.

\selectlanguage{french}%
\begin{table}[t]
\selectlanguage{english}%
\centering{}\caption{Comparison of theoretical expressions for the cross section ratio
for the two limit cases $\Omega\gg\left|\Delta\right|\gg\Gamma$ and
$\left|\Delta\right|\gg\Omega\gg\Gamma$.\foreignlanguage{french}{\label{tab:compa sect eff}}}
\begin{tabular*}{8.6cm}{@{\extracolsep{\fill}}cccc}
 &  &  & \tabularnewline
\hline 
\hline 
ref. & $\Omega\gg\left|\Delta\right|$ & $\Omega\ll\left|\Delta\right|$ & method\tabularnewline
\hline 
\multirow{2}{*}{\cite{steane1992}} & \multirow{2}{*}{1} & \multirow{2}{*}{2} & phenomenological calculation\tabularnewline[\doublerulesep]
 &  &  & \emph{two-level atom}\tabularnewline
\multirow{2}{*}{\cite{townsend1995}} & \multirow{2}{*}{${\displaystyle \frac{\Delta^{2}}{2\Gamma^{2}}}$} & \multirow{2}{*}{${\displaystyle \frac{3\Omega^{2}}{\Gamma^{2}}}$} & curve fitting / DAP\tabularnewline[\doublerulesep]
 &  &  & \emph{two-level atom}\tabularnewline
\multirow{2}{*}{\cite{pruvost2000}} & \multirow{2}{*}{${\displaystyle \frac{\Delta^{2}}{3\Gamma^{2}}}$} & \multirow{2}{*}{${\displaystyle \frac{\Omega^{2}}{2\Gamma^{2}}}$} & analytical expression / DAP\tabularnewline
\noalign{\vskip\doublerulesep}
 &  &  & \emph{two-level atom}\tabularnewline
\multirow{2}{*}{\cite{romain2011}} & \multirow{2}{*}{${\displaystyle \frac{\Delta^{2}}{6\Gamma^{2}}}$} & \multirow{2}{*}{${\displaystyle \frac{\Omega^{2}}{2\Gamma^{2}}}$} & analytical expression / DAP\tabularnewline
\noalign{\vskip\doublerulesep}
 &  &  & \emph{three-level atom}\tabularnewline
\hline 
\hline 
 &  &  & \tabularnewline
\end{tabular*}\selectlanguage{french}%
\end{table}

\selectlanguage{english}%

\section{Measurements}

Equations \ref{eq:A=00003Df(ellip)} and \ref{eq:A=00003Df(sigma)}
link the cross section ratio with the ellipticity. As ellipticity
can be measured experimentally, we have a way to determine experimentally
$\sigma_{R}$, and to compare this measurement to the theoretical
predictions. 

Our experimental set-up is described in \cite{distefano2004}: we
use a usual MOT in which each arm is formed by the retro-reflection
of an incident beam. To guarantee repeatability and reproducibility
of the measurement, some cares have to be taken. Indeed, the MOT laser
beams can create interference patterns on the cloud. So, its shape
is not symmetric and its density is not homogenous. Moreover, the
beam relative phase fluctuations make the patterns to move randomly
in time. To avoid these effects, we modulate the relative phases of
the beams at a frequency larger than 1 kHz. This modulation is faster
than the collective atomic response timescale. In addition, the shape
of the cloud is very sensitive to the alignment of the beams. We improve
the beam alignment to better than 0.1 mrad by observing the cloud
and adjusting its symmetry. 

In order to measure the ellipticity, we record the cloud fluorescence
with a CCD camera, the optical axis of which is perpendicular to the
coil axis. In that way, we obtain pictures showing a 2D projection
of the ellipsoid. The pictures are fitted on a 2D gaussian, giving
us the semi-axes $L_{\parallel}$ and $L_{\perp}$ of the ellipsoid.
A typical experimental measurement consists in recording the cloud
sizes as a function of $\Delta$, and repeating this sequence for
different values of $\Omega$. For each set of parameters, we record
10 pictures in order to improve the precision and to evaluate the
standard deviation.

Note that applying this method for very dilute or very dense clouds
leads to large measurement uncertainties. Indeed, this method requires
a good signal to noise ratio to monitor the cloud with a camera, which
is not the case for dilute clouds, i.e. for large detunings or low
laser intensities. On the other hand, in an asymmetric configuration,
a thick cloud is displaced by the shadow effect. Simple observations
show that the shadow effect degrades also the shape of the cloud,
typically when the laser is tuned close to resonance. This issue does
not exist for a MOT obtained with six independent beams.

\begin{figure}[t]
\begin{centering}
\subfloat{\selectlanguage{english}%
\centering{}\includegraphics[width=8.6cm]{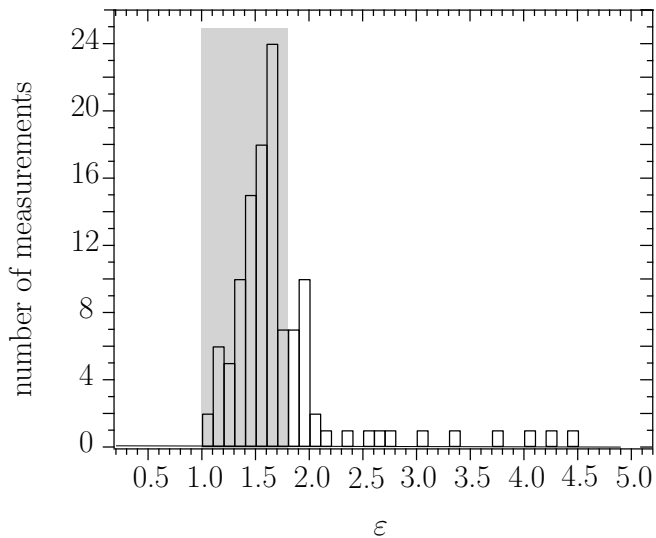}\selectlanguage{french}%
}
\par\end{centering}

\centering{}\caption{\selectlanguage{english}%
Measurements of the cloud ellipticity for 117 different values of
the MOT parameters. The theoretical range is shown in gray.\foreignlanguage{french}{\label{fig:hist}}\selectlanguage{french}%
}
\end{figure}

We first check the predicted range of ellipticity values. We measure
the cloud sizes for almost 120 different values of $\left(\Delta,\Omega\right)$.
Fig. \ref{fig:hist} shows all the measured values and how often they
have been measured. The result is in good agreement with the theoretical
range: about 90 \% (if we consider error bars) of the measured ellipticities
are inside the predicted range. No prolate cloud is observed. Furthermore,
most of the values over the limit correspond to pictures recorded
with a low signal to noise ratio, i.e. for which the measurement uncertainty
is probably underestimated.

\selectlanguage{french}%

\begin{figure}[t]
\selectlanguage{english}%
\centering{}\includegraphics[width=8.6cm]{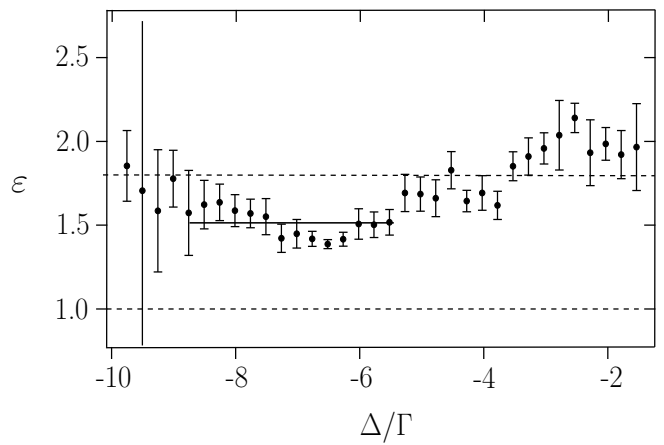}\foreignlanguage{french}{\caption{\selectlanguage{english}%
Evolution of the ellipticity $\varepsilon$ versus the laser detuning
$\Delta$ for $I_{+}\simeq3.6I_{sat}$. The dashed lines show the
theoretical limits of ellipticity values ($1<\varepsilon<1.81$).
The error bars represent the standard deviation from the 10 pictures.\label{fig:ellip_3.9mW}\selectlanguage{french}%
}
}\selectlanguage{french}%
\end{figure}

\selectlanguage{english}%
Fig. \ref{fig:ellip_3.9mW} shows the typical evolution of the ellipticity
as a function of the detuning. In this example, the minimum value
of the ellipticity is 1.4 and is measured around $\Delta=-6.5\Gamma$.
Points over the 1.81 limit are observed for $\left|\Delta\right|\lesssim3\Gamma$
, but as explained above, errors bars have probably been underestimated
for these measurements.

Strictly speaking, the two limit cases considered in Table \ref{tab:compa sect eff}
cannot be satisfied with our experimental parameter values. However,
for large enough $\Delta$ values, we approach the conditions $\left|\Delta\right|\gg\Omega\gg\Gamma$
. Moreover, as the measurement quality is poor for very large detunings,
we retain only intermediate values for our determination of $\varepsilon$.
For example, in the case of Fig. \ref{fig:ellip_3.9mW}, the estimation
of the cross section ratio is done in the range $-8.8\Gamma<\Delta<-5.5\Gamma$.
In this case, we have $\Omega\simeq\Gamma I_{+}/\sqrt{2}I_{sat}\simeq2.5\Gamma$
and $\left|\Delta\right|\gtrsim2\Omega$. Note that the secular limit
is satisfied because $\Omega^{2}+\Delta^{2}>35\Gamma^{2}$. In this
interval, the ellipticity is quite constant, as predicted, and we
get \foreignlanguage{french}{$\varepsilon=1.51\pm0.08$}, i.e. $\sigma_{R}=3.2\sigma_{L}$.
This value is in good agreement with \cite{pruvost2000,romain2011}
which gives the theoretical value $S_{th}$ of 3.1.

Table \ref{tab:compa sect eff dupli} summarizes the measurements
obtained in three different configurations. In each case, the measurement
and the predicted value of $S$ are very similar, with a relative
difference of less than 5 \%. Unfortunately, despite the good precision
on the ellipticity, the error on the cross section is significant.
This is due to the non-linearity of the relation between these two
quantities (Fig. \ref{fig:sect eff=00003Df(ellip)}). Moreover, we
get asymmetric errors because the derivative is also non-linear. As
a consequence the smaller the ellipticity, the better the precision.

\selectlanguage{french}%

\selectlanguage{english}%
\begin{table}[t]
\centering{}\caption{Comparison of the experimental determination of $S$ with the predictions.
For different intensity values, we give the averaged ellipticity $\varepsilon$
measured for a range of large detunings, the deduced cross section
ratio $S$ and the theoretical value $S_{th}$ from \cite{pruvost2000,romain2011}.\foreignlanguage{french}{\label{tab:compa sect eff dupli}}}
\begin{tabular*}{8.6cm}{@{\extracolsep{\fill}}ccccc}
 &  &  &  & \tabularnewline
\hline 
\hline 
\noalign{\vskip\doublerulesep}
$\Omega^{2}/\Gamma^{2}$ & detuning domain & $\varepsilon$ & $S$ & $S_{th}$\tabularnewline
\hline 
\noalign{\vskip\doublerulesep}
4.5 & $[-7.0\Gamma,-5.6\Gamma]$ & $1.39\pm0.05$ & ${\displaystyle 2.2\,_{-\,0.3}^{+\,0.3}}$ & 2.3\tabularnewline
\noalign{\vskip\doublerulesep}
7.0 & $[-8.8\Gamma,-5.5\Gamma]$  & $1.51\pm0.08$  & $3.2\,_{-\,0.7}^{+\,1.2}$ & 3.1\tabularnewline
\noalign{\vskip\doublerulesep}
9.4 & {[}$-9.2\Gamma,-5.5\Gamma]$ & $1.61\pm0.09$ & $4.9\,{}_{-\,1.6}^{+\,4.1}$ & 4.7\tabularnewline
\hline 
\hline 
 &  &  &  & \tabularnewline
\end{tabular*}
\end{table}

\begin{figure}[t]
\selectlanguage{english}%
\begin{centering}
\includegraphics[width=8.6cm]{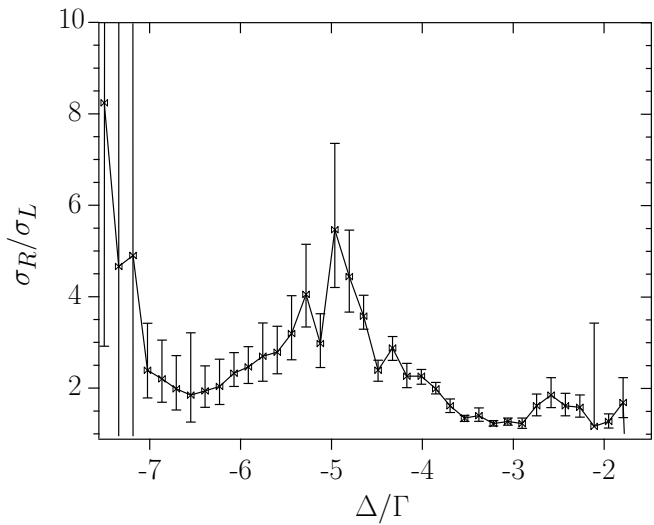}
\par\end{centering}

\selectlanguage{french}%
\caption{\selectlanguage{english}%
Evolution of the cross section ratio versus $\Delta$ for $I_{+}\simeq2.8I_{sat}$.\foreignlanguage{french}{\label{fig:sect eff=00003Df(delta)}}\selectlanguage{french}%
}
\end{figure}

Of course, the possibility to measure $\sigma_{R}$ is particularly
interesting for parameters where no theoretical predictions are available,
or where theoretical predictions are questionable because of the approximations.
It is in particular the case for smaller detunings and intensities.
Fig. \ref{fig:sect eff=00003Df(delta)} shows the evolution of the
cross section ratio for $-7.5\Gamma<\Delta<-1.8\Gamma$ and $I_{+}\simeq2.8I_{sat}$.
It is interesting to note that $S$ exhibits a maximum around $\Delta=-5\Gamma$,
with values larger than 5. These values are rather high compared to
those used in the literature but they are consistent with eq. 1. Close
to the resonance, the ratio varies between 1 and 2, values which lead
to a dense cloud as expected for these detunings. Whereas far from
resonance the ratio is a little bit larger than 2 but the density
is small due to a weak restoring.

\selectlanguage{french}%

\selectlanguage{english}%

\section{Conclusion}

In this paper, we propose a method to determine experimentally the
re-absorption cross section $\sigma_{R}$ in a cloud of cold atoms.
To our knowledge, it is the first method which allows to measure $\sigma_{R}$
for a large set of MOT parameter values. The method is non-destructive
and based on ellipticity measurements of the atomic cloud. The re-absorption
cross section can be measured very easily. The results of the measurements
are in good agreement with the calculation done with the dressed atom
picture in the limit studied here (large detunings and intermediate
intensities). We also make measurements for MOT parameters for which
no theoretical predictions have been done. The measurements show that
the cross section ratio is underestimated in the literature in this
case. The precision of this method can be as high as desired, it requires
only more acquisitions. A good precision is needed especially when
an important re-absorption is expected. Measuring the cross section
can be very useful to improve our description of the re-absorption
in future studies on spatio-temporal dynamics of the atomic cloud.

\end{document}